\documentclass[12pt,preprint]{aastex}

\usepackage{amsmath, amsthm, amssymb}
\usepackage{natbib}
\usepackage{epsfig}
\usepackage[update,prepend]{epstopdf}

\shorttitle{Kepler-76b: A hot Jupiter with superrotation evidence}
\shortauthors{Faigler et al.}

\begin{document}

\title{BEER analysis of {\it Kepler} and CoRoT light curves: \\
I. Discovery of Kepler-76b: A hot Jupiter with evidence for superrotation}


\author{S. Faigler\altaffilmark{1},  L. Tal-Or\altaffilmark{1}, T. Mazeh\altaffilmark{1},
            D. W. Latham\altaffilmark{2} and L. A. Buchhave\altaffilmark{3}}

\altaffiltext{1}{ School of Physics and Astronomy, Raymond and Beverly Sackler Faculty of Exact Sciences, Tel Aviv University, Tel Aviv  69978, Israel}

\altaffiltext{2}{ Harvard-Smithsonian Center for Astrophysics, 60 Garden St., Cambridge, MA 02138}

\altaffiltext{3}{ Niels Bohr Institute, University of Copenhagen, DK-2100 Copenhagen, Denmark}

\begin{abstract}
We present the first case in which the BEER algorithm identified a hot Jupiter in the {\it Kepler} light curve, and its reality was confirmed by orbital solutions based on follow-up spectroscopy. 
Kepler-76b was identified by the BEER algorithm, which detected the BEaming (sometimes called Doppler boosting) effect together with the Ellipsoidal and Reflection/emission modulations (BEER), at an orbital period of $1.54$ days, suggesting a planetary companion orbiting the $13.3$ mag F star. Further investigation revealed that this star appeared in the {\it Kepler} eclipsing binary catalog with estimated primary and secondary eclipse depths of $5\times 10^{-3}$ and $1\times 10^{-4}$ respectively.  Spectroscopic radial-velocity follow-up observations with TRES and SOPHIE confirmed Kepler-76b as a transiting $2.0 \pm 0.26$  $M_{\rm Jup}$ hot Jupiter.
The mass of a transiting planet can be estimated from either the beaming or the ellipsoidal amplitude. 
The ellipsoidal-based mass estimate of Kepler-76b is consistent with the spectroscopically measured mass while the beaming-based estimate is significantly inflated. We explain this apparent discrepancy as evidence for 
the superrotation phenomenon, which involves
eastward displacement of the hottest atmospheric spot of a tidally-locked planet 
by an equatorial super-rotating jet stream. 
This phenomenon was previously observed only for HD 189733b in the infrared. 
We show that a phase shift of $10.3 \pm 2.0$ degrees of the planet reflection/emission modulation, due to superrotation, explains the apparently inflated beaming modulation, resolving the ellipsoidal/beaming amplitude discrepancy.
Kepler-76b is one of very few confirmed planets in the {\it Kepler} light curves that show BEER modulations and the first to show superrotation evidence in the {\it Kepler} band. Its discovery illustrates for the first time the ability of the BEER algorithm to detect short-period planets and brown dwarfs. 
\end{abstract}



\section{Introduction}

CoRoT and {\it Kepler} have produced hundreds of thousands of
nearly uninterrupted high precision light curves
\citep{auvergne09, koch10} that enable detection of minute
astrophysical effects.
One of these is the the beaming effect, sometimes called Doppler boosting, induced by stellar radial velocity.
The effect causes a decrease (increase) of the brightness of any light source receding from (approaching) the observer \citep{rybicki79}, on the order of $4v_{\rm r}/c$, where $v_{\rm r}$ is the radial velocity of the source, and $c$ is the velocity of light. Thus, periodic variation of the stellar radial velocity due to an orbiting companion produces a periodic beaming modulation of the stellar flux.
\citet{loeb03} and \citet{zucker07} suggested to use this effect to identify {\it non-eclipsing} binaries and exoplanets in the light curves of CoRoT and {\it Kepler}. The precision of the two satellites is needed because even for short-period binaries, with large radial-velocity (RV) orbital amplitudes, the beaming effect
is small, on the order of 100--500  ppm (parts per million).

As predicted, several studies identified the beaming effect in short-period known eclipsing binaries \citep{vankerkwijk10,rowe11,carter11, kipping11, bloemen11,bloemen12, breton12, weiss12}. Yet, space missions data can be used to identify {\it non-eclipsing} binaries through detection of the beaming effect \citep{loeb03, zucker07}. However, the beaming modulation by itself might not be enough to identify a binary star, as periodic modulations could be produced by other effects, stellar variability in particular \citep[e.g.,][]{aigrain04}.

To overcome this problem, the BEER algorithm \citep{faigler11} searches for stars that show in their light curves a combination of the BEaming effect with two other effects that are produced by a short-period companion --- the Ellipsoidal and the Reflection modulations.
 The ellipsoidal variation \citep[e.g.,][]{morris85} is due to the tidal interaction between the two components (see a review by \citet{mazeh08}), while the reflection/heating variation (referred to herein as the reflection modulation) is caused by the luminosity of each component that falls on the facing half of
its companion \citep[e.g.,][]{wilson90,maxted02,harrison03,for10,reed10}.
Detecting the beaming effect together with the ellipsoidal and reflection periodic variations, with the expected relative amplitudes and phases, can indicate the presence of a small non-eclipsing companion.
Recently \citet{faigler12} reported RV confirmation of seven new non-eclipsing short-period binary systems in the {\it Kepler} field, with companion minimum masses in the range $0.07$--$0.4$ $M_{\odot}$, that were discovered by the BEER algorithm.

For brown-dwarfs or planetary companions the beaming effect is even smaller, on the order of 2--50 ppm. Interestingly, several studies were able to detect this minute effect in systems with {\it transiting brown dwarfs and planets} \citep{mazeh10, shporer11, mazeh12, jackson12, mislis12, barclay12}, indicating it may be possible to detect such {\it non-transiting} objects by identifying these effects in their host star light curves.

This paper presents the discovery of Kepler-76b, the first hot Jupiter detected by the BEER algorithm that was subsequently confirmed by TRES and SOPHIE RV spectroscopy.
It was identified by the BEER algorithm as a high-priority planetary candidate. Visual inspection of its light curve
revealed a V-shaped primary transit and a minute secondary eclipse, combined with beaming, ellipsoidal, and reflection amplitudes, consistent with a massive-planet companion. We noticed later that this star was listed in the {\it Kepler} eclipsing binary catalog \citep{prsa10,slawson11}. Based on this information spectroscopic follow-up observations were initiated for this target, which in turn confirmed its planetary nature.

Section~2 presents the BEER search and the initial analysis of the {\it Kepler} light curve,
Section~3 provides the details and results of the spectroscopic observations,
Section~4 describes the details and results of the light curve transits and occultations analysis,
Section~5 presents the detection of evidence for superrotation in the light curve
and Section~6 discusses the implications of, and conclusions from, the findings of this paper.

\section{The Photometric BEER search}
To identify candidates for low-mass companions we analyzed the {\it Kepler} raw light curves of the Q2 to Q10 quarters, spanning $831$ days. We visually identified $22$ time segments that showed instrumental artifacts in the photometry, and ignored data points within those segments, removing a total of $59.9$ days of data from the light curves. We also corrected two systematic jumps at {\it Kepler} times ($\rm BJD-2454833$) of $200.32$ and $246.19$ days. For each light curve, outliers were then removed by $4 \sigma$ clipping and detrending was performed using a cosine-transform filter, adapted to unevenly spaced data  \citep{mazeh10,mazeh12}, resulting in a cleaned and detrended light curve.
We then applied the BEER algorithm to
$41{,}782$ stars brighter than $13.7$ mag, with {\it Kepler} Input Catalog \citep{brown11} radius smaller than $4R_{\odot}$, calculating the Fast Fourier Transform (FFT) based power spectrum of the cleaned and detrended light curve of each star,  interpolated over the gaps.
Next, in order to avoid spurious peaks at long periods, we divided the full period range of each power spectrum into five sub-ranges: [0.3--1], [1--2], [2--5], [5--10], and [10--20] days, and identified the highest peak within each sub-range. For each of the five peaks we derived the BEER amplitudes and the estimated mass and albedo of the candidate companion \citep{faigler11}, assuming the peak corresponds to either the orbital period (beaming and reflection) or half the orbital period (ellipsoidal). The BEER amplitudes were calculated by fitting the data with the modified BEER model suggested by \citet{mazeh12}, that uses a Lambertian reflection/emission function.   

We then selected 26 candidates with the highest signal to noise ratio for the ellipsoidal and beaming amplitudes, and with estimated secondary mass smaller than $60$ $M_{\rm Jup}$ and implied albedo smaller than $0.5$.
One of these candidates was Kepler-76 ({\it Kepler} Input Catalog number 4570949), for which visual inspection revealed primary and secondary eclipses with depths of about $5\times 10^{-3}$ and $1\times 10^{-4}$, respectively. This candidate was also identified as a member of the {\it Kepler} Eclipsing Binary catalog \citep{prsa10}. 
Follow up spectroscopic observations confirmed the companion as a hot Jupiter. 
In forthcoming papers we will report on our observations of the 26 candidates, additional confirmation of a possible brown dwarf, and the false-positive rate of this sample.

\begin{figure*} 
\centering
\resizebox{18cm}{6.5cm}
{
\includegraphics{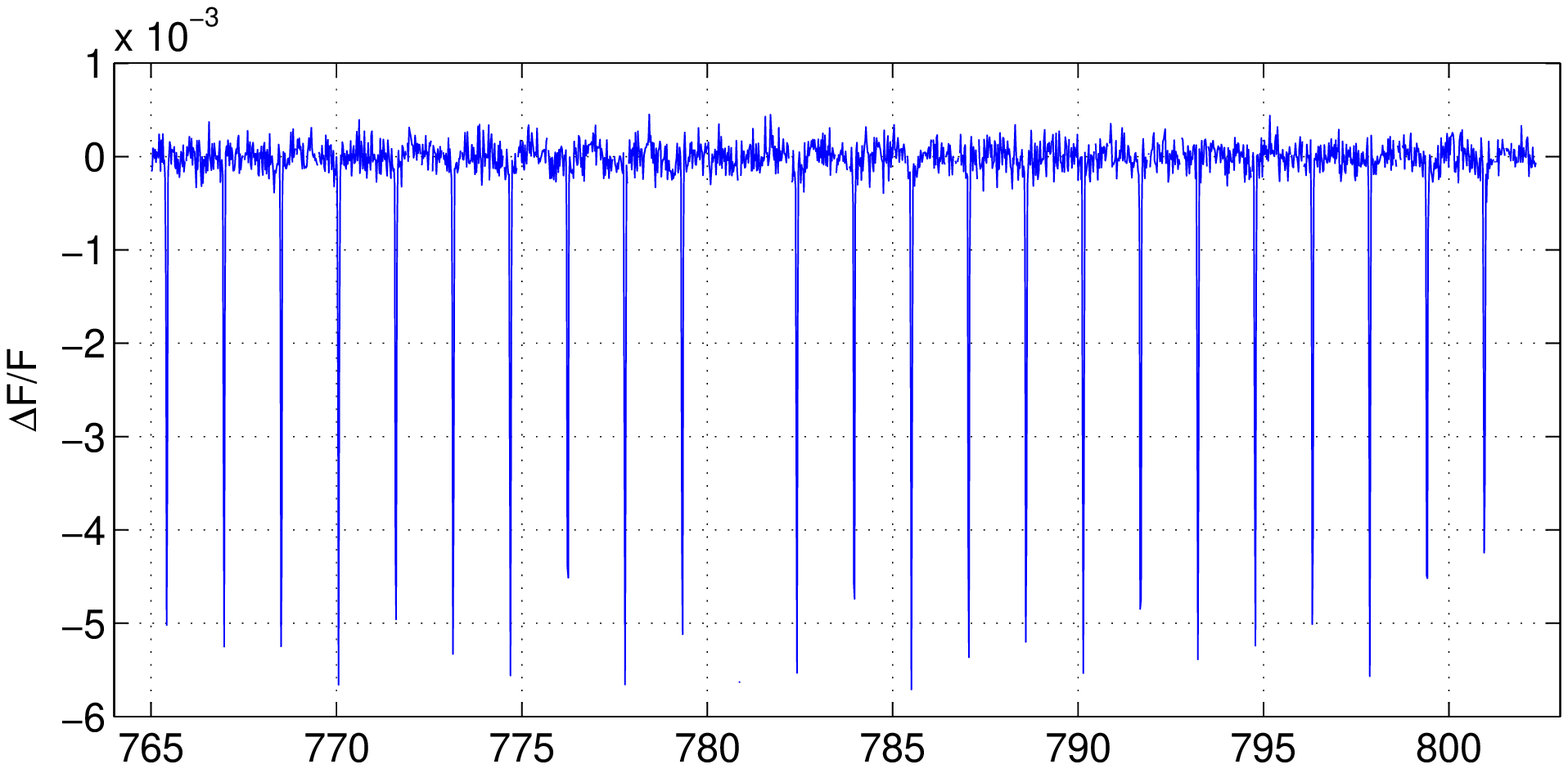}
}

\centering
\resizebox{18cm}{6.5cm} 
{
\includegraphics{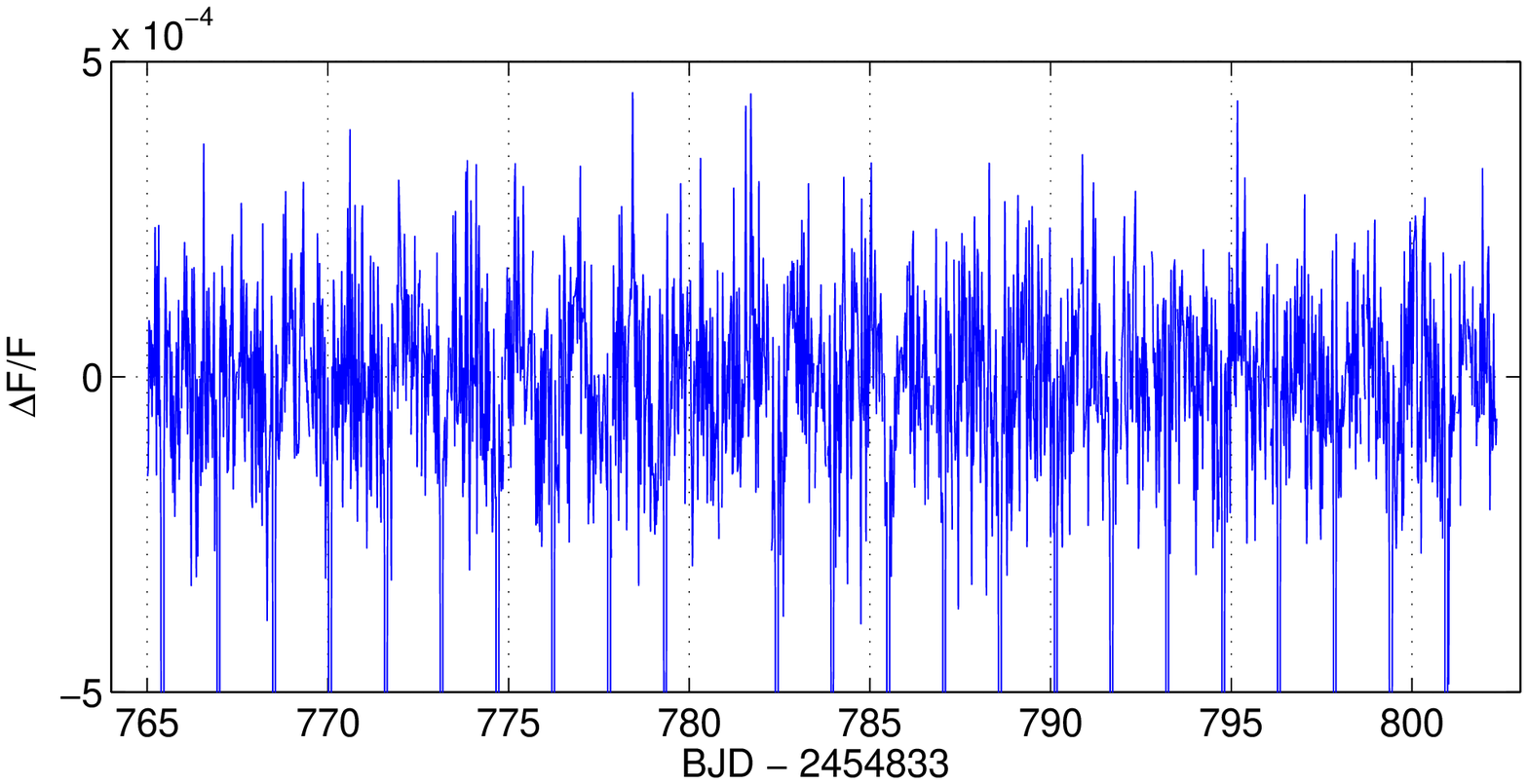}  
}

\caption{
The light curve of Kepler-76 for a selected time span of 37 days, after outlier removal and long-term detrending. 
Top: The untruncated light curve, showing the full depth of the transits. 
Bottom: The light curve with the core of the transit events truncated.
Note the different scales of the two plots. 
(The transit missing at time 780.9 fell in a short gap in the raw data.)
}
\label{figure:lc}
\end{figure*}

\begin{figure*} 

\centering
\resizebox{18cm}{8cm}
{
\includegraphics{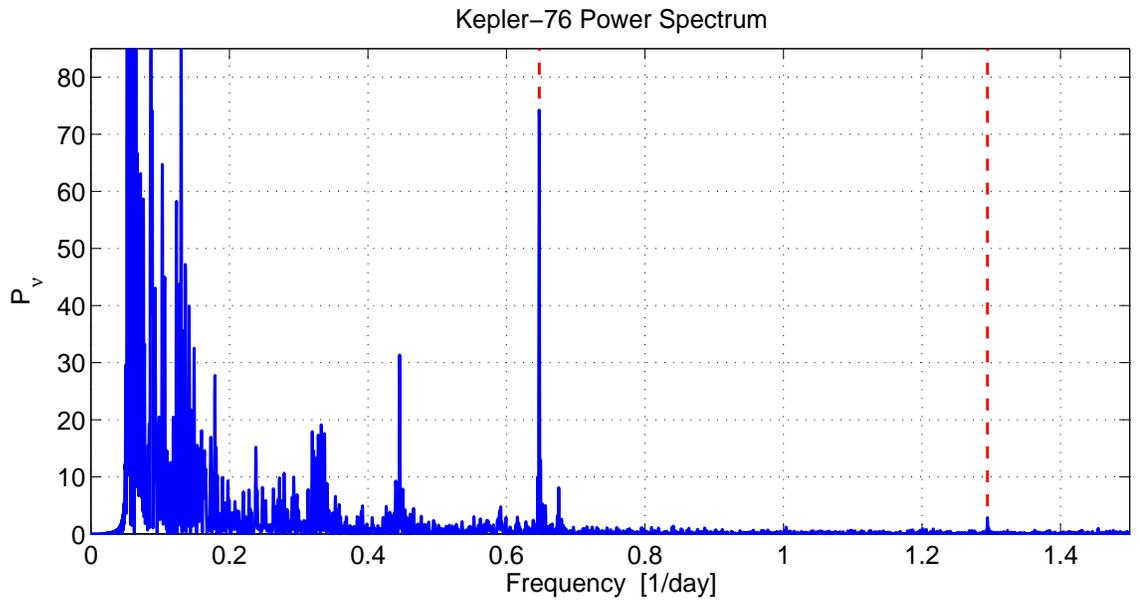}  
 }
\caption{
The FFT based power spectrum of the detection. 
The orbital period and half-orbital period peaks are marked by vertical dashed lines.
For clarity, only the frequency range of $0$--$1.5$ $\rm day^{-1}$ is plotted, since no significant peak was found for frequencies higher than $1.5$ $\rm day^{-1}$.
}
\label{figure:fft}
\end{figure*}

\begin{figure*} 

\centering
\resizebox{10cm}{9cm}
{
\includegraphics{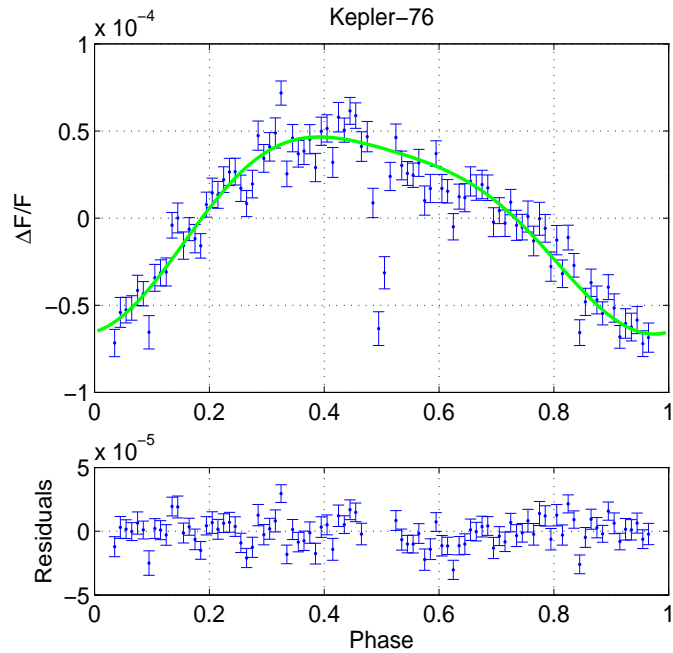}
}
\caption{
The folded cleaned light curve binned into 100 bins.  Phase zero is when the companion is closest to the observer, while phase $0.5$ is when the primary is closest to the observer, assuming a circular orbit. 
The error bars represent $1\sigma$ uncertainties of the median value of each bin, based on the scatter of data points within that bin.
The line presents the BEER model. The residuals of the data from the model are plotted in the bottom panel.
For clarity, the primary eclipse was removed. The secondary eclipse is clearly visible at phase 0.5 of the plot.  
}
\label{figure:fold}
\end{figure*}

We report here the BEER analysis results for the light curve of Kepler-76 after masking out the transit and occultation data points. We note that in our initial detection the BEER analysis was performed on the unmasked data, but the use a robust-fit function \citep{holland77} that identified the transit points as outliers resulted in similar measured BEER amplitudes. 
Table~\ref{table:kic} lists for Kepler-76 the coordinates and stellar properties from the Kepler Input Catalog \citep{brown11}, revised effective temperature estimate from \citet{pinsonneault12}, and the results of the BEER analysis. 
Figure~\ref{figure:lc} presents a short section of the `cleaned' \citep{mazeh10,faigler11} photometric data of the host star,  Figure~\ref{figure:fft} presents the FFT-based power spectrum, and Figure~\ref{figure:fold} shows the light curve folded with the detected period.
It is interesting to notice, by inspecting the cleaned light curves (Figure~\ref{figure:lc}) and the data and residuals r.m.s.~(Table~\ref{table:kic}), that the effects are significantly smaller than the light curve noise, to the point that the detected modulations almost can not be recognized by eye.
However, 
deriving the BEER photometric power spectrum from data with time spans of hundreds of days, produces a prominent detectable peak at the orbital period (Figure~\ref{figure:fft}).

%
\begin{deluxetable}{lrl}
\tabletypesize{\scriptsize}
\tablecaption{ Kepler-76: Stellar properties and BEER results}

\tablewidth{0pt}

\startdata
\tableline
RA  &  19:36:46.11 & Right ascension\\
DEC & 39:37:08.4   & Declination\\
$K_{\rm p}$$^a$ [mag] &13.3  & {\it Kepler} band magnitude\\
$T_{\rm eff}$$^a$ [K] & $6196$ & KIC Effective temperature  \\
$T_{\rm eff}$$^b$ [K] & $6409 \pm 95$ & Revised effective temperature  \\
$\log g$$^a$ [dex]  & $4.388$ & Surface gravity  \\
$[m/H]$$^a$ [dex] & $-0.033$ & Metallicity   \\
$R_*$$^a$ [$ R_{\odot}$] & $1.12$ & Primary radius  \\
$M_*$$^c$ [$ M_{\odot}$] & $1.12$ & Primary mass \\
$f_{\rm 3}$$^a$ & 0.056 & Average third light fraction \\

\tableline
BEER model: & & With transit and occultation points masked out\\

Period [days]    & $1.5449 \pm 0.0007$ & Orbital period\\

$T_{\rm 0}-2455000$$^d$ [BJD]   &   $737.49 \pm 0.19$  & Orbital zero phase ephemeris \\





Ellipsoidal [ppm]$^e$  & $21.5 \pm 1.7$ & Ellipsoidal semi-amplitude\\

Beaming [ppm]$^e$  & $15.6\pm 2.2$ & Beaming semi-amplitude \\

Reflection [ppm]$^e$ & $56.0 \pm 2.5$ & Lambertian reflection/emission semi-amplitude  \\

r.m.s.~cleaned data [ppm]  & $133$ & Root-mean-square of the cleaned and detrended data\\
r.m.s.~residuals [ppm]       & $127$ &  Root-mean-square of residuals of data from BEER model\\
\tableline

\enddata
\tablenotetext{\space}{ $^a$from Kepler Input Catalog}
\tablenotetext{\space}{ $^b$revised $T_{\rm eff}$ from \citet{pinsonneault12}}
\tablenotetext{\space}{ $^c$calculated from Kepler Input Catalog $\log g$ and $R$  }
\tablenotetext{\space}{ $^d$$T_{\rm 0}$ is the time in which the companion is closest to the observer, assuming a circular orbit}
\tablenotetext{\space}{ $^e$corrected for third light}
\label{table:kic}
\end{deluxetable}

\clearpage
\section{Spectroscopic observations}   %
Spectroscopic observations of the candidate were obtained between 29 May and 6 October 2012 with the Tillinghast Reflector Echelle Spectrograph \citep[TRES;][]{furesz08} mounted on the 1.5-m Tillinghast Reflector at the Fred Lawrence Whipple Observatory operated by the Smithsonian Astrophysical Observatory (SAO) on Mount Hopkins in Southern Arizona, using the medium resolution fiber at a spectral resolution of $44{,}000$, covering a spectral range from 385 to 910 nm. Exposures of a Thorium-Argon hollow-cathode lamp immediately before and after each exposure were used for wavelength calibration.  The spectra were extracted and rectified to intensity vs.\ wavelength using standard procedures developed by Lars Buchhave \citep{buchhave10}.

Additional spectroscopic observations were obtained between 17 July and 1 August 2012 with the SOPHIE spectrograph \citep{Perruchot08, bouchy09, bouchy13} mounted on the 1.93-m telescope at
Observatoire de Haute-Provence, France, using the High Efficiency mode ($R \sim 39,000$ at $550$\,nm) of the instrument. Spectra were extracted with the online standard pipeline.

Following a method similar to the Stellar Parameter Classification method  \citep[SPC,][]{buchhave12}, the atmospheric parameters of Kepler-76 were determined from the SOPHIE spectra by cross-correlating the observed spectral regions not affected by telluric lines against a library of synthetic spectra \citep{Hauschildt99}, with varying values of effective temperature $T_{\rm eff}$, surface gravity $\log g$, metallicity $[m/H]$ and rotational velocity $v \sin i$. 
For each of the observed spectra we derived the best set of parameters that yielded the highest correlation. This was done by fitting a second degree polynomial to the maximum correlation as function of each parameter around the synthetic spectrum that yielded the best correlation. The final parameter values for a star were taken as the mean of the parameter values derived for each observed spectrum of that star, weighted proportionally to the inverse of the scatter of the maximum around the fitted polynomial.

The Phoenix library of synthetic spectra we used spans the following intervals in atmospheric parameters: $3000 \rm K<T_{\rm eff}<10000 \rm K$, $-0.5<\log g<5.5$ (cgs), and $-1.5< [m/H]<+0.5$. The spacing in $T_{\rm eff}$ is $100 \rm K$ for $T_{\rm eff}<7000 \rm K$, and $200\rm K$ elsewhere. The spacing in $\log g$ and $[m/H]$ is $0.5$ dex. The interval and spacing of $v \sin i$ values in our algorithm are free parameters set by the user, since each synthetic spectrum chosen from the library is convolved with a rotational profile G(v) (e.g., \citet{gray05}, p. 465; \citet{santerne12}) and a Gaussian representing the instrumental broadening of the lines, just before calculating cross-correlation function.

The cross-correlation was performed using TODMOR \citep{zucker94, zucker03, zucker04} --- a two-dimensional correlation algorithm, assuming the light contribution of the secondary is negligible. In TODMOR, the cross-correlation functions are calculated separately for each echelle order, and then combined to a single cross-correlation function according the scheme proposed by \citet{zucker03}.
The atmospheric parameters found this way are listed in Table~\ref{table:spec}. The relatively large uncertainties result mainly from the addition of possible systematic errors \citep [see e.g.,][]{bruntt10,bruntt12,torres12}.

\begin{figure*} 
\centering
\resizebox{10cm}{8cm}
{
\includegraphics{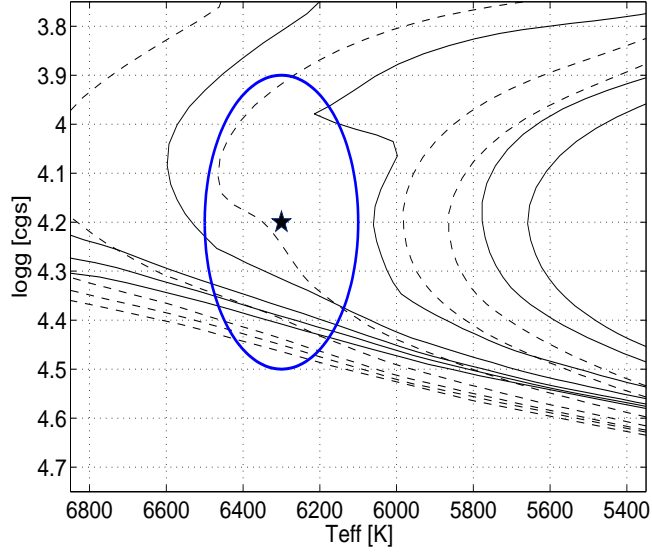}  
}
\caption{
Y$^2$ stellar isochrones, from \citet{demarque04}, of $0.2$--$10$ Gyr for metallicities  $[Fe/H] = 0.05$ (solid lines) and $[Fe/H] = -0.27$ (dashed lines). The estimated $T_{\rm eff}$ and $\log g$ of Kepler-76 with their uncertainties are marked by a star and an ellipse.
}
\label{figure:iso}
\end{figure*}

The primary mass was estimated using the atmospheric parameters derived from the spectra and a grid of Y$^2$ stellar isochrones \citep{yi01, demarque04}. This was done by taking into account all age and mass values that fall into the ellipsoid in the $(T_{\rm eff}, \log g, [Fe/H])$ space defined by the atmospheric parameters and their errors. To illustrate the process Figure~\ref{figure:iso} shows two sets of Y$^2$ stellar isochrones of $0.2, 0.4, 1, 2, 4, 8,$ and $10$ Gyr --- one for $[Fe/H]=0.05$ (solid lines) and one 
for $[Fe/H]=-0.27$ (dashed lines). The ellipse defined by the estimated $T_{\rm eff}$ and $\log g$ and their uncertainties is also shown. A lower limit of 0.2 Gyr on the stellar age was set to ignore possible pre-main sequence solutions. This procedure yielded a mass estimate of $1.20 \pm 0.09 M_{\odot}$. Following \citet{basu12} we have conservatively doubled the mass errors to take into account possible uncertainties in stellar model parameters.

Radial velocities were derived for the TRES observations in two different
ways, as described in detail by \citet{faigler12}.  First, we
derived absolute velocities using cross-correlations of the observed
spectra against the template from our library of synthetic spectra
that yielded the best match (with $T_{\rm eff} = 6000 \rm K$, $\log g = 4.0$ cgs,
$v \sin{i} = 12 \,{\rm km\,s^{-1}}$ and solar metallicity).  The absolute
velocity analysis used just the spectral order containing the MgIb triplet
and was calibrated using IAU RV standard stars.  With the goal of achieving
better precision we also derived velocities using about two dozen spectral
orders, correlating the individual observations against a template based on
the strongest exposure.  Thus the multi-order velocities are relative to
the observation chosen as the template.  They are reported in Table~\ref{table:RV}.

For the SOPHIE observations, radial velocities were derived by computing the weighted cross-correlation function (CCF) of the spectra with a numerical spectral mask of a G2V star \citep{baranne96, pepe02}.
For the last five exposures, which were contaminated by scattered moon light, we subtracted the sky using the fiber B spectrum \citep{santerne09}, before deriving the radial velocities.  
Table~\ref{table:RV} lists the radial-velocity measurements and their uncertainties.

%
\begin{table}
\caption{Atmospheric parameters of Kepler-76}

\small \begin{tabular}{lr}
\\
\hline
$T_{\rm eff}$ [K] & $6300 \pm 200$  \\
$\log g$ [dex] & $4.2 \pm 0.3$ \\
{[m/H]} [dex] & $-0.1 \pm 0.2 $ \\
$v \sin i$ [km s$^{-1}$] & $6.5 \pm 2$ \\
$M_*$ [$M_{\odot}$] & $1.2 \pm 0.2$ \\
\hline
\end{tabular}
\label{table:spec}
\end{table}

\begin{table}
\caption{ Radial-velocity measurements}
\tiny \begin{tabular}{lrrc}
\\
\hline
  Time [BJD$-2456000$]   & RV [km s$^{-1}$] & $\sigma$ [km s$^{-1}$]  & Instrument \\ 
\hline
76.930366 & 0.581    &     0.069  &  TRES \\
83.895818 & 0.161    &     0.110 & TRES \\
84.868623 & 0.546     &    0.072 & TRES \\
87.836880	 &  0.607  &       0.103 & TRES \\
107.916759 &   0.586     &    0.098 &  TRES \\
115.796005 &  0.727    &     0.114 & TRES \\
117.775394 & 0.100     &    0.082 & TRES \\
207.682884      &                   0    &       0.069 & TRES \\

126.378842          &        -4.999         & 0.036  & SOPHIE \\
128.575439           &      -5.597         & 0.036 & SOPHIE \\
129.562543           &        -5.081         & 0.056 & SOPHIE \\
130.416149          &         -5.560         & 0.091 & SOPHIE \\
131.389038          &        -5.341     &    0.061  & SOPHIE \\
137.536798         &          -5.196    &     0.132 & SOPHIE \\
138.488304         &          -5.194      &   0.057 & SOPHIE \\
139.470933         &          -5.615    &     0.080 & SOPHIE \\
140.470516         &          -4.992    &     0.114 & SOPHIE \\
141.450747         &          -5.171    &     0.082 & SOPHIE \\
\hline
\end{tabular}
\label{table:RV}
\end{table}

\clearpage

\begin{figure*} 
\centering
\resizebox{16cm}{8cm}
{
\includegraphics{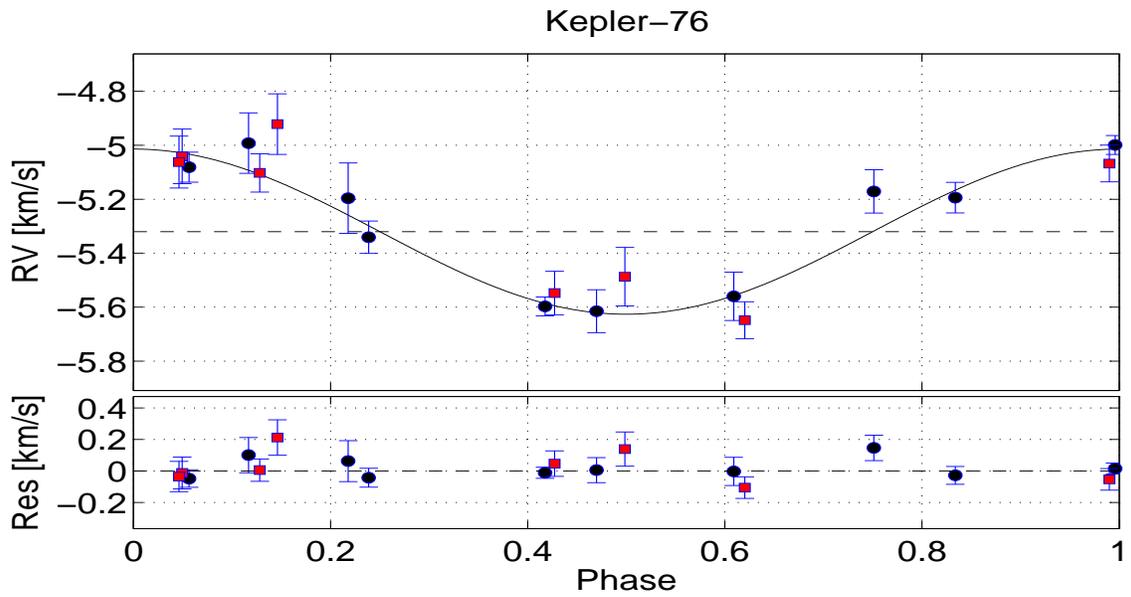}  
}
\caption{
The RV measurements folded at the derived orbital period. 
In the top panel, the solid line presents the photometry-constrained orbital RV model and the horizontal-dashed line indicates the center-of-mass velocity. Circles denote SOPHIE RV points and squares denote TRES RV points.
The residuals are plotted at the bottom panel.  Note the different scales of the upper and lower panels. 
}
\label{figure:RV}
\end{figure*}

The first RV measurements of Kepler-76 showed variability consistent with the photometric orbital phase, so we continued observations in order to allow an orbital solution independent of the BEER analysis. 
The derived eccentricity of the solution was statistically indistinguishable from zero, so we reran the solution with eccentricity fixed to zero.
Figure~\ref{figure:RV} shows the follow-up RV measurements and the velocity curve for the orbital solution, folded with the period found, and the top section of Table~\ref{table:RVsol} lists the derived orbital elements for the independent RV solution.
The center-of-mass velocities $\gamma_{\rm T}$ and $\gamma_{\rm S}$ for the independent
RV sets from TRES and SOPHIE differ by 5.68 ${\rm km\,s^{-1}}$.  This is
because the TRES velocities are relative to the strongest observation,
while the SOPHIE velocities are meant to be on an absolute scale.  If the
absolute TRES velocities derived using the Mg b order are used instead of
the relative velocities, $\gamma_{\rm T} = -5.18 \,{\rm km\,s^{-1}}$, quite close
to the SOPHIE value of $\gamma_{\rm S} = -5.31 \,{\rm km\,s^{-1}}$.  For the joint
analysis reported below, the two independent velocity sets were shifted to
a common zeropoint using the $\gamma$ velocities reported in Table~\ref{table:RVsol}.

Next, in order to obtain a combined solution from photometry and RV measurements, we reran the RV model using the photometric period and ephemeris, with their uncertainties, as  priors. The bottom section of Table~\ref{table:RVsol} lists the orbital elements derived from this photometry-constrained RV solution, and the estimated minimum secondary mass, $M_{\rm p} \sin i$.

\clearpage

%
%
\begin{deluxetable}{lrl}
\tabletypesize{\scriptsize}
\tablecaption{Orbital elements based on RV measurements} 

\tablewidth{0pt}
\startdata
\tableline
\\
$ N$  & 18  & Number of RV measurements \\
span  [days]   &  $130.8$   & Time span of measurements\\

\\
\tableline
\\

Independent RV solution: \\

$T_{\rm 0}-2456000$ $^a$  [BJD]&  $126.76 \pm 0.03$   & Orbital ephemeris  \\ 

$P$ [days] & $1.5474 \pm 0.0021$  & Orbital period \\

$\gamma_{\rm T}$  [km s$^{-1}$]& $0.365 \pm 0.042$  & Zero point velocity of TRES\\

$\gamma_{\rm S}$  [km s$^{-1}$]& $-5.315 \pm 0.019$  & Zero point velocity of SOPHIE\\

$K_{\rm \mathrm\scriptscriptstyle RV}$  [km s$^{-1}$] & $0.308 \pm 0.020$  & Radial velocity semi-amplitude \\

\\
\tableline
\\

RV solution constrained by the  \\
photometric period and ephemeris:\\
$T_{\rm 0}-2456000$ $^a$  [BJD]&  $126.77 \pm 0.023$   & Orbital ephemeris  \\ 

$P$ [days] & $1.5450 \pm 0.0005$  & Orbital period \\

$\gamma_{\rm T}$  [km s$^{-1}$]& $0.328 \pm 0.032$  & Zero point velocity of TRES\\

$\gamma_{\rm S}$  [km s$^{-1}$]& $-5.320 \pm 0.020$ & Zero point velocity of SOPHIE\\

$K_{\rm \mathrm\scriptscriptstyle RV}$  [km s$^{-1}$] & $0.306 \pm 0.020$  & Radial velocity semi-amplitude \\
\\
$M_{\rm p} \sin i$ [$M_{\rm \mathrm Jup}$] &  $1.96 \pm 0.25$   & Minimum companion mass       \\

\\

\enddata
\tablenotetext{\space}{$^a$  $T_{\rm 0}$ is the time in which the companion is closest to the observer }
\label{table:RVsol}
\end{deluxetable}

\clearpage

\section{Photometric modeling of the light curve}
For a more complete photometric analysis of this transiting hot Jupiter we used the {\it Kepler} light curves of the Q2 to Q13 quarters, spanning $1104$ days. First we fitted the cleaned and detrended data with the BEER model while masking out data points in or around the transits and occultations. The fitted amplitudes, after correction for a third light using the KIC estimate, are listed in Table~\ref{table:SR}. We then subtracted the BEER model from the data and analyzed the data points in and around the transits and occultations.
For that we ran a Markov chain Monte Carlo (MCMC) analysis, while fitting the transit data points using a long-cadence integrated \citet{mandel02} model with quadratic limb darkening, assuming a circular orbit. The model limb darkening coefficients could not be constrained, so we kept them fixed at values interpolated from \citet{claret11} using the stellar parameters derived from spectroscopy. We then fitted the occultation data keeping the geometric parameters derived from the transit fixed, and assuming a linear limb darkening coefficient of $0.5$ for the planet, while looking for the occultation depth that best fits the data.

\begin{figure*} 
\centering
\resizebox{17cm}{8cm}
{
\includegraphics{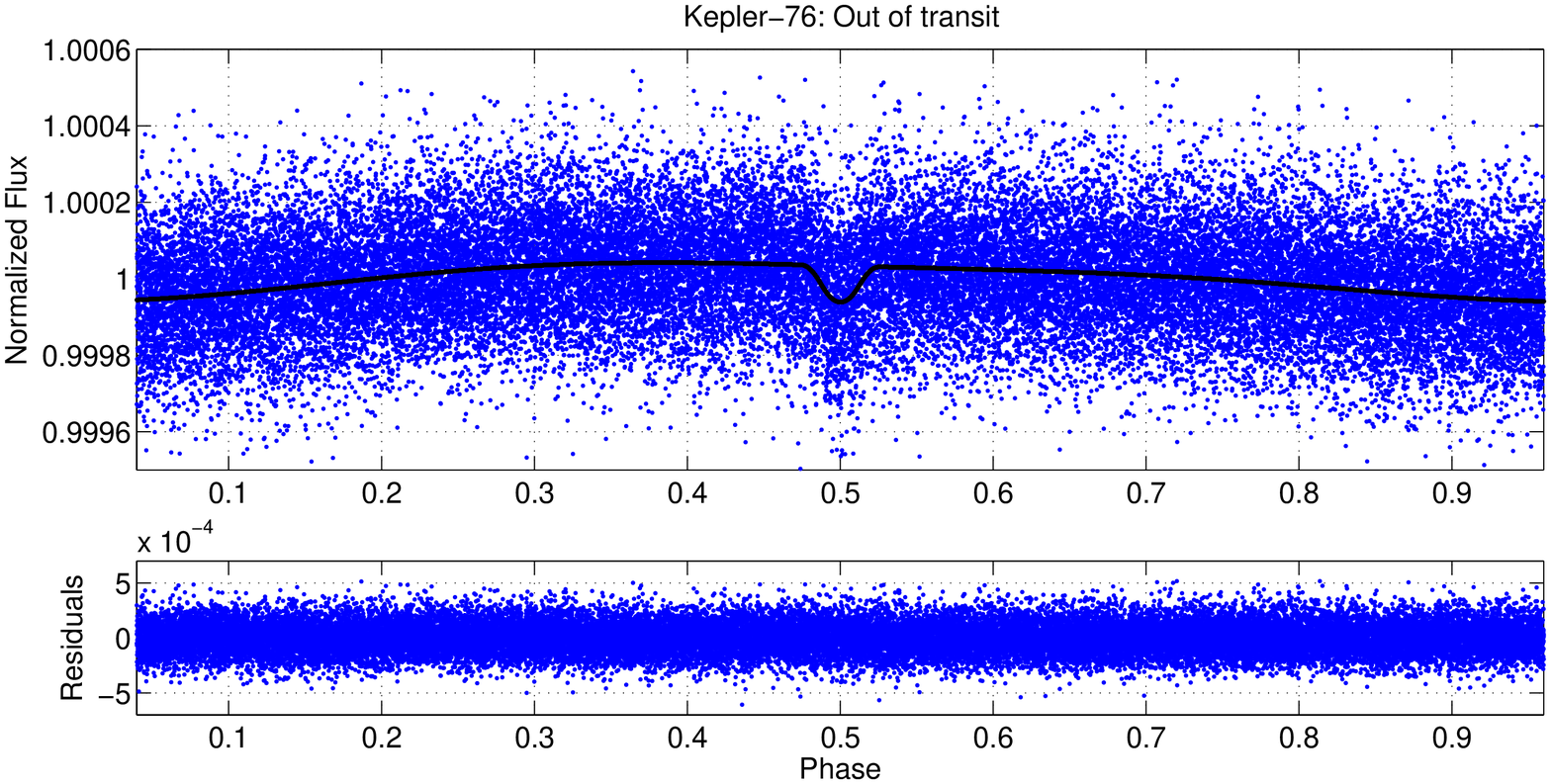}  
}
\centering
\resizebox{17cm}{8cm}
{
\includegraphics{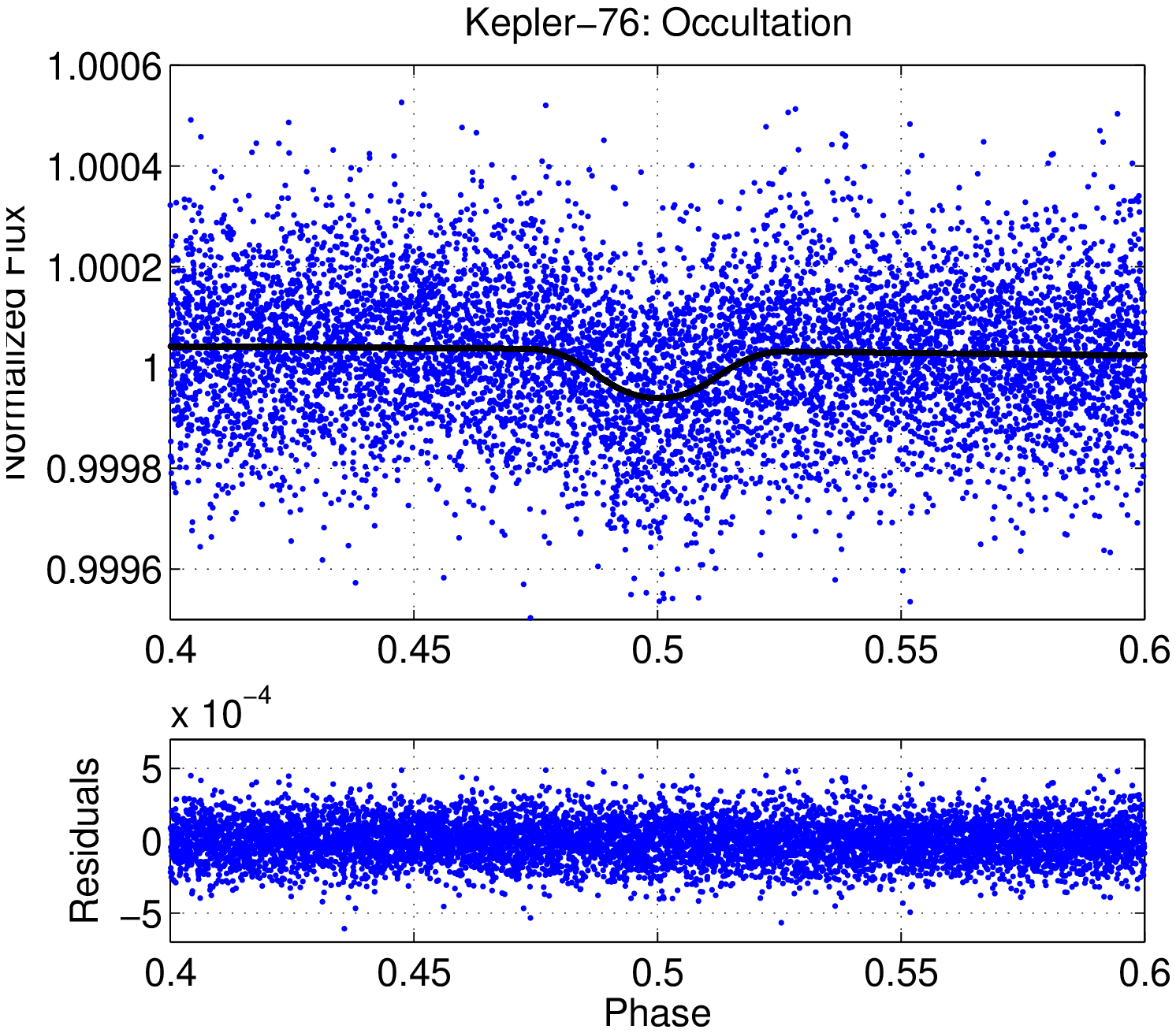}  
\includegraphics{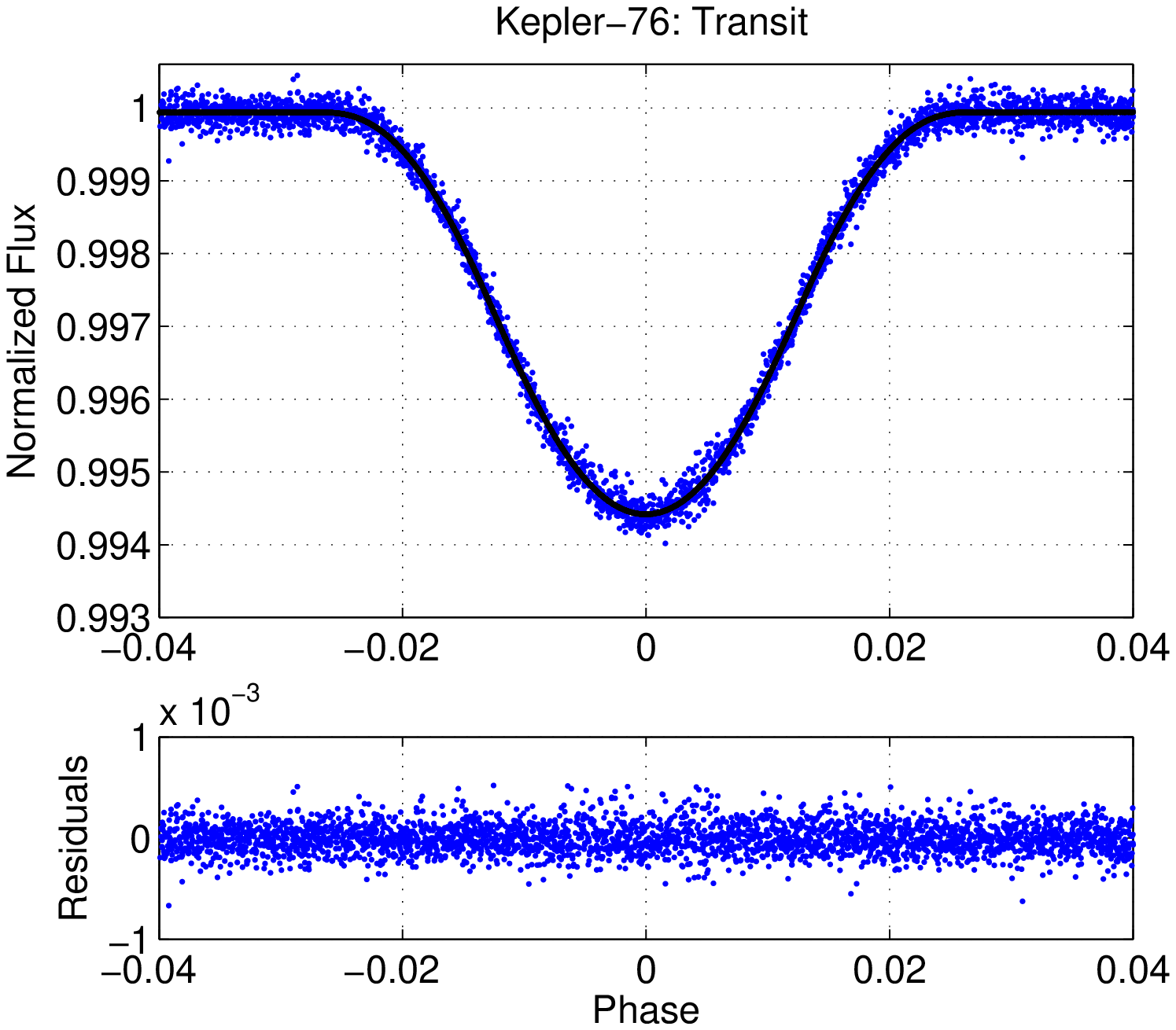}  
}
\caption{
Kepler-76 cleaned and detrended data points and best-fit model, both folded at the orbital period. In the top panel of each plot, the solid line presents the best-fit model, and the dots present the data points.
The residuals are plotted in the bottom panel.  Note the different scales of the upper and lower panel of each plot. 
}
\label{figure:trans}
\end{figure*}

Figure~\ref{figure:trans} presents the cleaned and detrended data points and the best-fit \citet{mandel02}  model combined with the BEER model, both folded at the orbital period. 
Table~\ref{table:MCMC} lists the priors and the MCMC medians and $1 \sigma$ uncertainties of the posterior parameters. 
We note here that the orbital period $P$ and time of primary transit $T_0$ listed in Table~\ref{table:MCMC} were derived from the transit data points, thus yielding high-accuracy estimates. The same parameters, listed in Table~\ref{table:kic}, were derived from the out-of-transit data, thus yielded much lower-accuracy estimates.
The $T_0$ values listed in both tables were not corrected for the {\it Kepler} timing error 
(http://archive.stsci.edu/kepler/timing\_error.html), which should be taken in account when comparing them to non-{\it Kepler} observations.

\begin{deluxetable}{lrl}
\tabletypesize{\scriptsize}
\tablecaption{
Photometric light curve transit and occultation model}
\tablewidth{0pt}
\tablehead{
\colhead{Parameter}  & \colhead{Median and $1 \sigma$ uncertainty}
}

\startdata

{\bf Priors:} \\
$M_*$ [$M_{\odot}$] & $ 1.2 \pm 0.2 $ & Primary mass \\  
$a_*$  & $ 0.313 $ & Star linear limb darkening coefficient \\  
$b_*$  & $ 0.304 $ & Star quadratic limb darkening coefficient \\  
$a_{\rm p}$ & $ 0.5 $ & Planet linear limb darkening coefficient\\  
$b_{\rm p}$ & $ 0 $ & Planet quadratic limb darkening coefficient\\  
$f_{\rm 3}$ & $ 0.056 $ & Average third light fraction (From KIC) \\  

\tableline
\\
{\bf Posteriors:} \\
$T_{\rm 0}$ [BJD]  &  $ 2454966.54811 \pm 0.00007$   & Time of primary transit \\  
Period [days]   & $1.54492875  \pm 0.00000027 $   &  Orbital period\\     
$i$ [deg] & $ 78.0 \pm 0.2 $  & Orbital inclination \\  
$R_*/a$  & $ 0.221 \pm 0.003 $ & Fractional primary radius\\ 
$R_{\rm p}/a$  & $ 0.0214 \pm 0.0008 $ & Fractional planet radius \\  

$d_{\rm 2}$ [ppm] & $98.9 \pm 7.1$ & Occultation depth \\
\tableline
\\
 
$\rho_*/\rho_{\odot}$  & $ 0.522 \pm 0.019 $ &  Primary density relative to the Sun density \\  
$R_*$ [$R_{\odot}$]  &  $1.32 \pm 0.08$ & Primary radius\\  
$R_{\rm p}$ [$R_{\rm Jup}$]  &  $1.25 \pm 0.08$ & Planet radius\\  
$b$  & $ 0.944 \pm 0.011 $ & Impact parameter \\
 

\enddata
\label{table:MCMC}
\end{deluxetable}

\clearpage

\section{Inflated beaming amplitude and planet equatorial super-rotating jet}        %
The spectroscopic RV observations and the light curve transit and occultation analysis yielded independent orbital solutions with ephemeris and period nicely consistent with the BEER ephemeris and period. 
To compare the measured RV amplitude with the beaming-based predicted RV amplitude, one needs to evaluate the $\alpha_{\rm beam}$ factor, which corrects for the Doppler shift of the stellar spectrum relative to the observed band of the telescope \citep{rybicki79,faigler11}.

To estimate the $\alpha_{\rm beam}$ value, we used spectra from the library of \citet{castelli04} models close to the estimated temperature, metallicity and gravity of the primary star, numerically shifting them relative to the {\it Kepler} response function, while taking into account the photon counting nature of {\it Kepler}  \citep{loeb03, bloemen11, faigler12}. 
For clarity we note that by definition
$\alpha_{\rm beam}=\frac{3-\alpha}{4}=\frac{\langle B \rangle}{4}$, 
where $\alpha$ is the power-law index used by \citet{loeb03} and $\langle B \rangle$ is the photon weighted bandpass-integrated beaming factor used by \citet{bloemen11}.
The result of this calculation gave  $\alpha_{\rm beam}=0.92\pm 0.04$, resulting in an RV semi-amplitude of $K_{\rm beam}=1.11 \pm 0.17 \, {\rm km \, s^{-1}}$.
The RV semi-amplitude predicted from the beaming effect was $3.5$ times larger than the measured amplitude, with a difference significance of about $4.5\sigma$ between the two. 

A possible explanation for this inflated photometric beaming amplitude might be 
a phase shift of the reflection signal, due to 
the superrotation phenomenon, which involves
eastward advection of gas by an equatorial super-rotating jet within the atmosphere of a co-rotating companion.
 \citet{showman02} predicted through a 3D atmospheric circulation model that {tidally-locked}, short period planets develop a fast eastward, or {\it superrotating}, jet stream that extends from the equator to latitudes of typically $20^\circ-60^\circ$. They showed that in some cases (depending on the imposed stellar heating and other factors) this jet causes an eastward displacement of the hottest regions by $10^\circ-60^\circ$ longitude from the substellar point, resulting in a phase shift of the thermal emission phase curve of the planet. This prediction was confirmed by \citet{knutson07,knutson09} through {\it Spitzer} infrared observations of HD 189733, which indicated a phase shift of $16^\circ \pm 6^\circ$ in the $8\ \mu {\rm m}$ band and $20^\circ-30^\circ$ in the $24\ \mu {\rm m}$ band.

In general, what we call a reflection modulation is actually the light scattered off the planet in combination with radiation absorbed and later thermally re-emitted at different wavelengths.
The two processes are controlled by the Bond albedo, $0<A_{\rm B}<1$, and the day--night heat redistribution efficiency, $0<\epsilon<1$, which can be constrained only if observations of the phase modulation or the secondary eclipse are available in different wavelengths \citep{cowan11}. This makes it impossible to distinguish between reflected and re-radiated photons from the single band {\it Kepler} light curve we have in hand. \citet{cowan11} discuss HAT-P-7 as an example, and show that its {\it Kepler} light curve can be explained as mostly reflected light at one limit, to mostly thermal emission at the other limit, with an entire range of models between them being consistent with the light curve. This is important for the current discussion, as we expect superrotation to shift only the thermal re-emission, while leaving the scattered light component unshifted. 

To estimate the maximum fraction of the reflection amplitude originating from thermal re-emission in our case,  we follow \citet{cowan11} and estimate the no albedo, no redistribution, effective day side temperature $T_{\rm \epsilon=0}\approx 2670 K$, which translates in the {\it Kepler} band to a maximum reflection amplitude $A_{\rm ref} \approx 37 \ {\rm ppm}$. This means that the measured amplitude of $\approx 50 \ {\rm ppm}$ can be explained mostly by thermal re-emission. The actual fraction of thermal emission in this case is probably smaller, but this calculation illustrates that the fraction of thermal emission in the visual {\it Kepler} light curve phase modulation may be significant, making it a worthy effort to look for a superrotation phase shift in the light curve.

We suggest here, that if such a phase shift is present in the {\it Kepler} light curve, it will show up in our phase curve model mainly as an inflated beaming amplitude. 
To illustrate that we consider a simple superrotation model consisting of a phase-shifted geometric reflection/emission combined with a beaming modulation,
\begin{equation}
\mathcal{M}_{SR}=-A_{\rm ref}\cos (\phi + \delta_{SR}) + A_{\rm beam}\sin \phi
= -A_{\rm ref}\cos \delta_{SR} \cos \phi + (A_{\rm beam}+\underline{A_{\rm ref} \sin \delta_{SR}}) \sin \phi  \ ,  
\end{equation}
where $A_{\rm ref}$ is the reflection/emission semi-amplitude, $A_{\rm beam}$ is the beaming semi-amplitude, $\phi$ is the orbital phase relative to mid-transit, and $\delta_{SR}$ is the superrotation phase shift angle. This model suggests that if a phase shift is present {\it and} the reflection amplitude is larger than, or of the order of, the beaming amplitude, the underlined term in Eq.~1 may add substantially to the amplitude of the $\sin \phi$ modulation, mimicking an inflated beaming effect.
 
To test our conjecture that the beaming/ellipsoidal inconsistency is a result of a superrotation phase shift of the reflection/emission phase modulation, 
we fitted the data using the derived system parameters (Tables~\ref{table:MCMC} and \ref{table:spec}) and the BEER effects equations \citep{faigler11}, while looking for the planetary mass, geometric albedo, and phase shift of the Lambertian phase function that minimized the $\chi^2$ of the fit. 
Adding the phase shift parameter to the model resulted in a decrease of the $\chi^2$ value by $90$, relative to the no-phase-shift model, indicating a substantially better agreement of the data with a model that combines beaming,  ellipsoidal and a phase shifted Lambertian reflection. An F-test shows that fitting the data while allowing for a phase shift, as opposed to the no-phase-shift null model, yields a better fit with a confidence level better than $9\sigma$.
Table~\ref{table:SR} lists the amplitudes derived by the BEER analysis, the planetary mass derived directly from the beaming versus the ellipsoidal amplitudes, and the spectroscopic RV derived planetary mass. 
The table then lists the planetary mass, phase-shift angle and geometric albedo resulting from the superrotation model. 
The superrotation phase shift estimate is small and well within the theoretical limit of $60^0$ predicted by \citet{showman02}. In addition, the derived planetary mass estimate 
is well within the $1\sigma$ range of the RV measured planetary mass, indicating that assuming superrotation resolves the inconsistency and provides a good estimate for the planetary mass, derived {\it solely} from the {\it Kepler} photometry, given a good stellar model.

%
\begin{deluxetable}{lrl}
\tabletypesize{\scriptsize}
\tablecaption{BEER amplitudes, planetary mass, and superrotation phase shift angle}
\tablewidth{0pt}
\tablehead{
}
\startdata

BEER model:\\

Ellipsoidal  [ppm]$^a$& $21.1 \pm 1.7$ & Ellipsoidal semi-amplitude\\

Beaming [ppm]$^a$   & $13.5\pm 2.0$ & Beaming semi-amplitude\\

Reflection [ppm]$^a$  & $50.4 \pm 2.0$ & Reflection semi-amplitude\\   

$a_{\rm 2s}$  [ppm]$^a$ & $-2.7 \pm 1.0$ & $\sin 2\phi$ coefficient\\  
\\

\tableline
\\
$M_{\rm p,beam}$ [$ M_{\rm Jup}$] & $7.2 \pm 1.4$ & Planetary mass derived from beaming     \\
$M_{\rm p,ellip}$ [$ M_{\rm Jup}$] & $2.1 \pm 0.4$  & Planetary mass derived from ellipsoidal          \\
$M_{\rm p,RV}$ [$ M_{\rm Jup}$] & $2.00 \pm 0.26$   & Planetary mass measured by spectroscopic RV          \\

\\
\tableline
\\
Superrotation phase-shift model solution:\\
$M_{\rm p,SR}$ [$ M_{\rm Jup}$] & $2.1 \pm 0.4$   & Planetary mass        \\
$\delta_{\rm SR}$ [deg] & $10.3 \pm 2.0$      & Superrotation phase shift angle       \\
$A_{\rm g}$  & $0.23 \pm 0.02$      & Geometric albedo       \\

$ N $        & 31468 &   Number of data points \\
$\chi^2$ & 31465 &  $\chi^2$ of a model allowing for a phase shift \\
$\chi^2_{\rm null}$ & 31555 &  $\chi^2$ of a zero phase shift model (null model) \\
\\
\enddata
\tablenotetext{\space}{$^a$corrected for third light}
\label{table:SR}
\end{deluxetable}

\clearpage

\section{Discussion}

This paper presents a new hot-Jupiter companion, Kepler-76b, initially identified by the BEER algorithm, and later confirmed by spectroscopic observations. The BEER detection was based on the photometrically measured amplitudes of the BEaming, Ellipsoidal and Reflection effects, that were consistent with a planetary companion. This is just the third confirmed planet in the {\it Kepler} field, after HAT-P-7b \citep{welsh10} and TrES-2b \citep{barclay12}, that its host light curve exhibits the three phase curve effects, and is the faintest of the three stars. It is also one of a few confirmed grazing exoplanets, showing a V-shaped transit and a partial occultation. 

We have identified an inconsistency between the beaming amplitude and the spectroscopically measured radial velocity. Similar inconsistencies  between the planetary mass derived from the beaming amplitudes and the mass derived from the ellipsoidal amplitude were noticed  previously by several authors for KOI-13 \citep{mazeh12,shporer11} and TrES-2 \citep{barclay12}. We suggest here that these inconsistencies can be explained by a phase shift of the planetary thermal modulation due to the equatorial superrotation phenomena predicted by \citet{showman02} and later observed by \citet{knutson07,knutson09} in the infrared for HD 189733. In such cases we should be able to measure the superrotation phase-shift angle from the visual band {\it Kepler} light curve of the system. As we do not expect scattered light to exhibit such a phase shift, visual band detection of the superrotation phase shift may yield a constraint on the ratio of scattered light to thermally re-emitted light from the planet.

\begin{figure*} 


\centering
\resizebox{18cm}{11cm}
{
\includegraphics{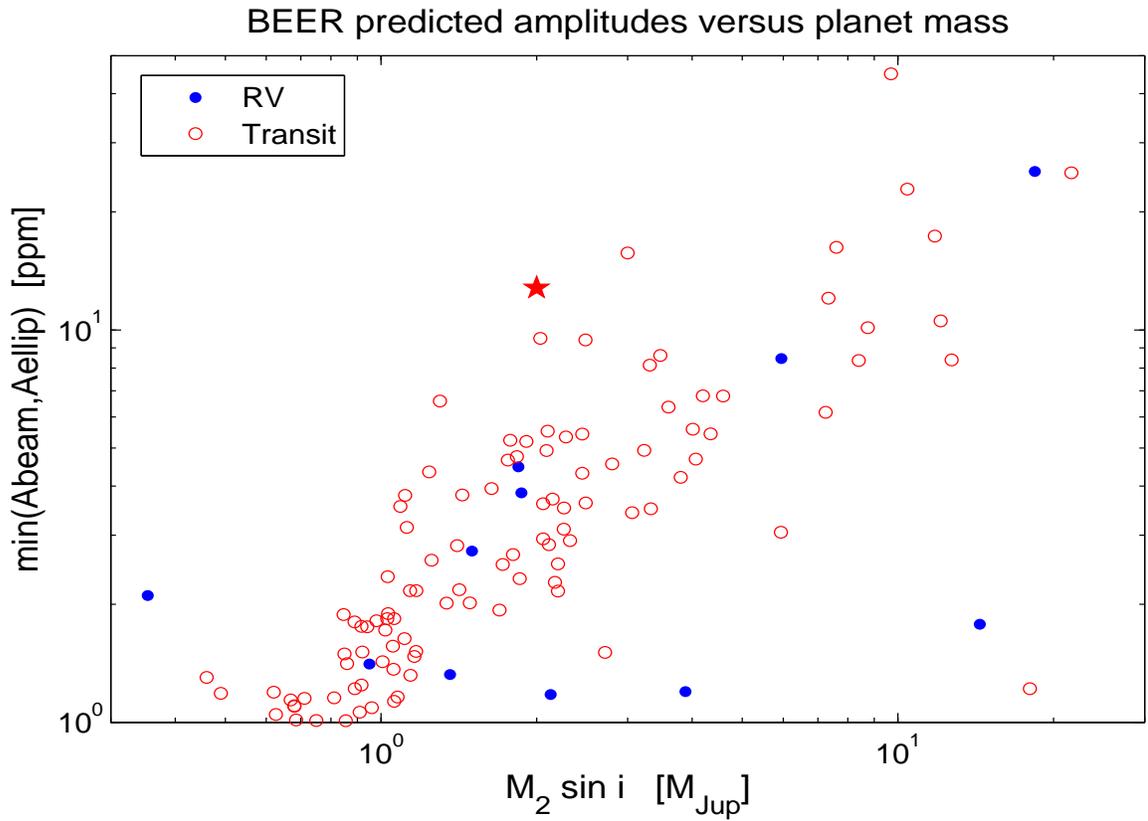}  
}

\caption{
The minimum of the predicted beaming and ellipsoidal semi-amplitudes, as a function of the secondary mass, for known exoplanets of mass higher than $0.5 M_{\rm Jup}$ and period shorter than 30 days. Blue dots mark exoplanets detected by RV, red circles mark exoplanets detected by transit, and the red star marks the measured amplitude of Kepler-76b. 
}
\label{fig:sens}
\end{figure*}

Finally, we wish to briefly comment on the sensitivity of the BEER algorithm. The detection presented here exhibits the lowest-mass companion identified so far by the algorithm, indicating its possible current detection limit. As we require that a BEER candidate must show statistically significant beaming {\it and} ellipsoidal effects to be considered valid, we choose here the minimum of the semi-amplitudes of the two effects as the BEER detectability parameter of a planet. To estimate our ability to detect more planets and brown dwarfs, Figure~\ref{fig:sens} presents the calculated value of this parameter, using the \cite{faigler11} equations, for known exoplanets of mass higher than $0.5 M_{\rm Jup}$ and period shorter than 30 days, as of January 2013 (http://exoplanet.eu/), together with the measured value for Kepler-76b. 
The figure shows that there are six transiting and one RV detected planets with calculated amplitudes higher than that of Kepler-76b, suggesting that these systems could have been detected by the BEER algorithm, if their stellar and instrumental noise were similar to that of  Kepler-76.  
It is also apparent that further enhancement of the algorithm sensitivity could significantly increase the number of potentially detectable planets. Given the fact that BEER can detect similar {\it non-transiting} objects, we expect to find more objects once we improve our detection threshold. 

\clearpage

\acknowledgments
We are indebted to Shay Zucker for numerous helpful discussions, and to Ignasi Ribas for discussing the thermal phase shift. 
We thank the referee Steven Bloemen for his valuable remarks and suggestions, and especially for his comment about the $\alpha_{beam}$ calculation.
The research leading to these results has received funding from the European Research Council under the EU's Seventh Framework Programme (FP7/(2007-2013)/ ERC Grant Agreement No.~291352).
This research was supported by the ISRAEL SCIENCE FOUNDATION (grant No.~1423/11).
We feel deeply indebted to the team of the Kepler mission, that enabled us to search and analyze their unprecedentedly accurate photometric data.
All the photometric data presented in this paper were obtained from the 
Multimission Archive at the Space Telescope Science Institute (MAST). 
STScI is operated by the Association of Universities for Research in 
Astronomy, Inc., under NASA contract NAS5-26555. Support for MAST for 
non-HST data is provided by the NASA Office of Space Science via grant 
NNX09AF08G and by other grants and contracts.
We thank the Kepler mission for partial support of the spectroscopic
observations under NASA Cooperative Agreement NNX11AB99A with the
Smithsonian Astrophysical Observatory, DWL PI.
We are indebted to Andrew H. Szentgyorgyi, who led the TRES project, and to
Gabor F\H{u}r\'{e}sz for his many contributions to the success of the
instrument. We thank Robert P. Stefanik, Perry Berlind, Gilbert A.
Esquerdo, and Michael L. Calkins for obtaining the TRES observations, and
Allyson Bieryla and Jessica Mink for help with the data analysis.
This paper is based in part on observations made at Observatoire de Haute Provence (CNRS), France. We are grateful to the OHP director and team for the allocation of the SOPHIE observing time. We are also thankful for the help of the night assistants, that enabled us to obtain the spectra presented here. OHP observations were supported by the OPTICON network. OPTICON has received research funding from the European Community's Seventh Framework Programme. 

{\it Facilities:} 
\facility{FLWO:1.5m(TRES)} , \facility{OHP:1.93m(SOPHIE)}

\end{document}